\documentclass{emulateapj}

\shorttitle{Strong lensing by subhalos}
\shortauthors{Zackrisson et al.}

\begin{document}

\title{Strong lensing by subhalos in the dwarf-galaxy mass range I: Image separations}
\author{E. Zackrisson\altaffilmark{1,2,3}$^*$, T. Riehm\altaffilmark{2}, O. M\"oller\altaffilmark{4}, K. Wiik\altaffilmark{1}, \& P. Nurmi\altaffilmark{1}}
\altaffiltext{*}{E-mail: ez@astro.uu.se}
\altaffiltext{1}{Tuorla Observatory, University of Turku, V\"ais\"al\"antie 20, FI-21500 Piikki\"o, Finland}
\altaffiltext{2}{Stockholm Observatory, AlbaNova University Center, 106 91 Stockholm, Sweden}
\altaffiltext{3}{Department of Astronomy and Space Physics, Box 515, 751 20 Uppsala, Sweden}
\altaffiltext{4}{Schloss-Wolfsbrunnenweg 66, 69118 Heidelberg, Germany}

\begin{abstract}
The cold dark matter scenario predicts that a large number of dark subhalos should be located within the halo of each Milky-way sized galaxy. One tell-tale signature of such dark subhalos could be additional milliarcsecond-scale image splitting of quasars previously known to be multiply-imaged on arcsecond scales. Here, we estimate the image separations for the subhalo density profiles favoured by recent N-body simulations, and compare these to the angular resolution of both existing and upcoming observational facilities. We find, that the image separations produced are very sensitive to the exact subhalo density profile assumed, but in all cases considerably smaller than previous estimates based on the premise that subhalos can be approximated by singular isothermal spheres. Only the most optimistic subhalo models produce image separations that would be detectable with current technology, and many models produce image separations that will remain unresolved with all telescopes expected to become available in the foreseeable future. Detections of dark subhalos through image-splitting effects will therefore be far more challenging than currently believed, albeit not necessarily impossible.   

\end{abstract}





\section{Introduction} 
While the cold dark matter (CDM) scenario has been very successful in explaining the formation of large-scale structures in the Universe \citep[see e.g.][for a review]{Primack}, its predictions on the scales of individual galaxies have not yet been confirmed in any convincing way. One particularly interesting feature of the CDM model is the high level of halo substructure generated. According to current simulations, each galaxy-mass CDM halo should contain large numbers of subhalos \citep[typically accounting for $\approx 5$--10\% of its total mass; e.g.][]{Nurmi et al.,Diemand et al. a} in the dwarf-galaxy mass range. However, these subhalos do not appear to correspond to luminous structures, as the dark halo of the Milky Way would then contain a factor of 10--100 more satellite galaxies than observed, provided that each subhalo corresponds to a luminous dwarf galaxy \citep[e.g.][]{Klypin et al.}. A similar lack of dwarf galaxies compared to the number of dark halos predicted is also evident on scales of galaxy groups \citep{Tully et al.}. One way out of this problem is to assume that most of these low-mass halos correspond to so-called dark galaxies \citep[][]{Verde et al.}, i.e. objects of dwarf-galaxy mass which either do not contain baryons or in which the baryons have not formed many stars. While a number of very faint satellite galaxies have recently been detected \citep[e.g.][]{Zucker et al.,Simon et al.}, it is still far from clear that these exist in sufficent numbers to account for the subhalos predicted by CDM \citep[][]{Simon & Geha}.

Gravitational lensing may in principle offer a route to detecting completely dark galaxies, and the magnification associated with dark halo substructures has long been suspected to be the cause of the flux ratio anomalies observed in multiply-imaged quasars \citep[e.g.][]{Mao & Schneider,Kochanek & Dalal}. However, the subhalos predicted by current CDM simulations appear too few to explain the flux ratios of many of these systems \citep[e.g.][]{Mao et al.,Metcalf a,Maccio & Miranda}. This indicates that other mechanisms, like lensing by low-mass field halos outside the main lens \citep{Metcalf b,Miranda & Maccio}, stellar microlensing in the lens galaxy \citep{Schechter & Wambsganss} and absorption or scintillation in the interstellar medium \citep{Mittal et al.} may also be at work. 
 
A similar but more direct way of putting the CDM subhalo predictions to the test could be to search for small-scale {\it image separations} produced by strong lensing from subhalos. \citet{Yonehara et al.} argue that by targeting quasars that are already known to be multiply-imaged on arcsecond scales, there should be a significant probability of detecting image splitting by subhalos at scales of milliarcseconds (so-called millilensing or sometimes mesolensing). If the sources are extended and can be resolved at scales smaller than the Einstein radii of the subhalos, then not only the number densities of the subhalos but also their internal density profiles can be constrained \citep{Inoue & Chiba b}. It has been argued that this technique may become observationally feasible with ALMA \citep{Inoue & Chiba a} or future space-VLBI missions like VSOP-2 \citep{Inoue & Chiba c}. If correct, this would constitute a major step forward in the study of dark halo substructures.

These proposals do however rely on the assumption that the image separations (or Einstein radii) of subhalo lenses are similar to those produced by singular isothermal spheres. This is an assumption that is difficult to justify, since neither simulations, theoretical arguments nor observations favour a density profile of this form for dark matter halos in the relevant mass range. 

N-body simulations based on the CDM paradigm typically predict dark matter halos to have inner density profiles of the form $\rho(r)\propto r^{-\alpha}$ with central density slopes $\alpha\approx 1$ \citep[e.g.][hereafter NFW]{NFW}, whereas the singular isothermal sphere (SIS) assumes $\alpha=2$. While the SIS has proved to be a successful model for observed galaxy-mass lenses \citep[e.g.][]{Rusin et al.,Koopmans et al.}, this is believed to be due to the luminous baryons residing in the inner regions of these objects. This component contributes substantially to the overall mass density in the centre, and its formation over cosmological time scales may also have caused the CDM halo itself to contract, thereby steepening the inner slope of its density profile \citep[e.g.][]{Gnedin et al.,Maccio et al.,Gustafsson et al.,Kampakoglou}.

Dwarf galaxies are well-known to be more dark matter dominated than their more massive counterparts, and some of the mechanisms proposed for keeping the missing satellites of the Local Group dark assume that most of their baryons have been lost early in the history of the Universe, or are prevented from collapsing into their central regions \citep[e.g.][]{Barkana & Loeb,Read et al.}. Hence, it seems reasonable that the inner slope $\alpha$ of their density profiles should be shallower than that of an SIS. Indeed, \citet{Ma} finds that to explain the distribution of angular separations among strong lenses, the lens population must shift from an SIS-type profile ($\alpha = 2$) to something more resembling the NFW profile ($\alpha \approx 1$) at halo masses of $\lesssim 10^{12}\ M_\odot$. Observational constraints on the slope of the innermost density profile of both field and satellite galaxies in the dwarf galaxy mass-range typically favour a slope even shallower than this \citep[$\alpha \approx 0$--1; e.g.][]{de Blok & Bosma,Spekkens et al.,Zackrisson et al.,Gilmore et al.}. This is potentially bad news for the detectability of low-mass subhalos through millilensing effects, since the image separation expected from strong lensing by a halo of fixed virial mass is extremely sensitive to the slope of its inner density profile. In fact, as pointed out by \citet{Trentham et al.}, the image separation expected from strong lensing by dark galaxies would be unobservably small if their density profiles are of NFW type. An inner slope of the density profile even shallower than this would just make the problem worse. 

As low-mass halos in the field fall into the potential wells of larger halos, tidal interactions with their host halos and with other subhalos will cause these objects to lose mass. Since this mass loss preferentially takes place in the outer regions of the subhalos, their density profiles change over time \citep[e.g.][]{Hayashi et al., Kazantzidis et al.}, thereby altering their lensing properties compared to more isolated field halos of the same mass. Here, we take a critical look at the prospects for strong-lensing detections of dark subhalos in the dwarf-galaxy mass range, by deriving the image separations expected for subhalo density profiles favoured by current simulations. 

\section{Dark halo density profiles}
When deriving the subhalo density profiles and the corresponding image separations, we assume a $\Lambda$CDM cosmology with $\Omega_\Lambda=0.7$, $\Omega_\mathrm{M}=0.3$ and $h=0.7$ ($H_0=100h$ km s$^{-1}$ Mpc$^{-1}$). As we are mainly interested in image-splitting of the individual macroimages of multiply-imaged quasars, we compute all density profiles at a typical lens redshift of $z_\mathrm{l}=0.5$ and adopt a source redshift of $z_\mathrm{s}=2.0$.  The impact of of other choices for $z_\mathrm{l}$ and $z_\mathrm{s}$ on our results are discussed in section 3. We moreover assume that halos contain no material beyond their virial radius $r_\mathrm{vir}(z_\mathrm{l})$, defined as the radius at which the mean enclosed density equals $\Delta_\mathrm{vir}$ times the critical density of the Universe at redshift $z_\mathrm{l}$. Here, $\Delta_\mathrm{vir}$ is calculated using approximations from \citet{Bryan & Norman}.

The majority of papers on the density profiles of CDM halos have so far focused on relatively isolated halos. When such objects are accreted by more massive halos and become subhalos, substantial mass loss occurs, preferentially from their outer regions. Because of this, some sort of modification of these density profiles is in order when applying them to the study of subhalos. However, since some authors have previously modelled the lensing properties of subhalos using the original, unstripped profiles, we will here consider such models as well, so that the the importance of the stripping on the resulting image separations becomes clear. It is moreover possible that the original profiles may be reasonable approximations for recently accreted objects or low-mass halos which are spatially correlated with the parent halo but formally located outside its virial radius \citep[e.g.][]{Nurmi et al.,Diemand et al. b}. Even though the latter objects are located far away from the central regions of their host halos, they are predicted to exist in great numbers and may end up projected close to the images of strongly-lensed quasars.

\subsection{Unstripped profiles}
The original NFW density profile, relevant for relatively isolated CDM halos, is given by: 
\begin{equation}
\rho_\mathrm{NFW}(r)=\frac{\rho_\mathrm{i}}{(r/r_\mathrm{S,NFW})(1+r/r_\mathrm{S,NFW})^{2}}, 
\end{equation}
where $r_\mathrm{S,NFW}$ is the characteristic scale radius of the halo.

Because of the persisting controversies concerning the most appropriate fitting formula for the density profiles of CDM halos, we also consider the \citet[hereafter M99]{Moore et al.} profile, which -- while having an inner density profile steeper than that of the original NFW formula -- gives a fit of comparable quality to simulation data for low-mass halos \citep{Navarro et al.}:
\begin{equation}
\rho_\mathrm{M99}(r)=\frac{\rho_0}{(r/r_\mathrm{S,M99})^{1.5}\left[1+(r/r_\mathrm{S,M99})^{1.5}\right]}.
\end{equation}
\citet[][hereafter N04]{Navarro et al.} themselves do however advocate yet another profile, 
\begin{equation}
\rho_\mathrm{N04}(r)=\rho_{S,N04}\exp (-2/\beta \left[ (r/r_{S,N04})^\beta -1  \right] ),
\end{equation}
with the property of asymptotically approaching a finite central density. Here, $\rho_{S,N04}$ is the density at $r_\mathrm{S,N04}$, i.e. $\rho_\mathrm{i}/4$, and $\beta\approx0.17$ for halos in the dwarf-galaxy mass range. 

To derive the $R_\mathrm{S}$ parameters of the NFW, M99 and N04 profiles, we compute these in units of the radius $r_\mathrm{-2}$ at which the power-law slope of the density profile obeys $\mathrm{d}\ln \rho /\mathrm{d}\ln r = -2$. This quantity is linked to the virial radius $r_\mathrm{vir}$ through the concentration parameter $c=r_\mathrm{vir}/r_\mathrm{-2}$, with a median value of \citep{Bullock et al.}: 
\begin{equation}
c=\frac{9}{1+z_\mathrm{l}}\left(\frac{M h}{1.5\times 10^{13}M_\odot}\right)^{-0.13}.
\end{equation}
 Contrary to many other $c(M,z)$ models, this one has been demonstrated to extend into the dwarf-galaxy mass range \citep{Colin et al.}.
Once the $r_\mathrm{S}$ parameters are determined, $\rho_\mathrm{i}$, $\rho_\mathrm{S}$ and $\rho_\mathrm{0}$ can then be derived by requiring that the correct mass is contained within $r_\mathrm{vir}$.

In contrast to these CDM-motivated halo models, the commonly used SIS model has a density profile of the form:
\begin{equation}
\rho_\mathrm{SIS}=\frac{\sigma_\mathrm{v}^2}{2\pi G r^2},
\end{equation}
where $\sigma_\mathrm{v}$ is the line-of-sight velocity dispersion.

\subsection{Stripped profiles}
Based on N-body simulations of subhalos orbiting within a static and spherical host potential, \citet[][hereafter H03]{Hayashi et al.} suggested a subhalo density profile that represents a simple modification of the NFW profile of its progenitor: 
\begin{equation}
\rho_\mathrm{H03}(r)=\frac{f_\mathrm{t}}{1+(r/r_\mathrm{te})^3}\rho_\mathrm{NFW}(r),
\label{H03eq}
\end{equation}
where $f_\mathrm{t}$, the central density reduction factor, and $r_\mathrm{te}$, the effective tidal radius of the subhalo (in units of the progenitor scale radius $r_\mathrm{S}$), are both related to the mass fraction of the subhalo that remains bound, $m_\mathrm{bnd}$, through relations given in H03. For simplicity, we calculate the properties of the subhalo progenitor profile by assigning it an approximate mass $M_\mathrm{prog}=M_\mathrm{sub}/m_\mathrm{bnd}$.
\begin{figure*}
\plottwo{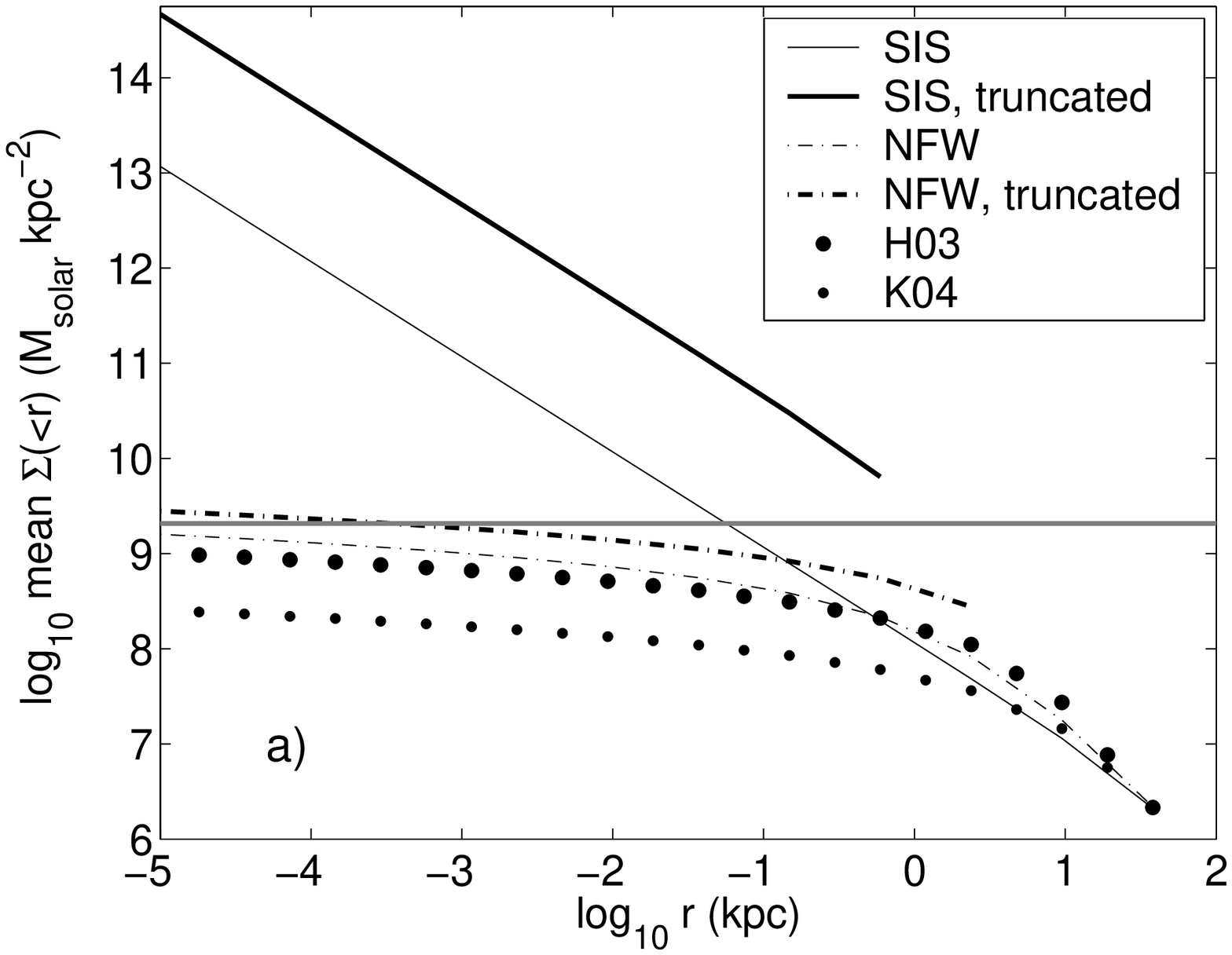}{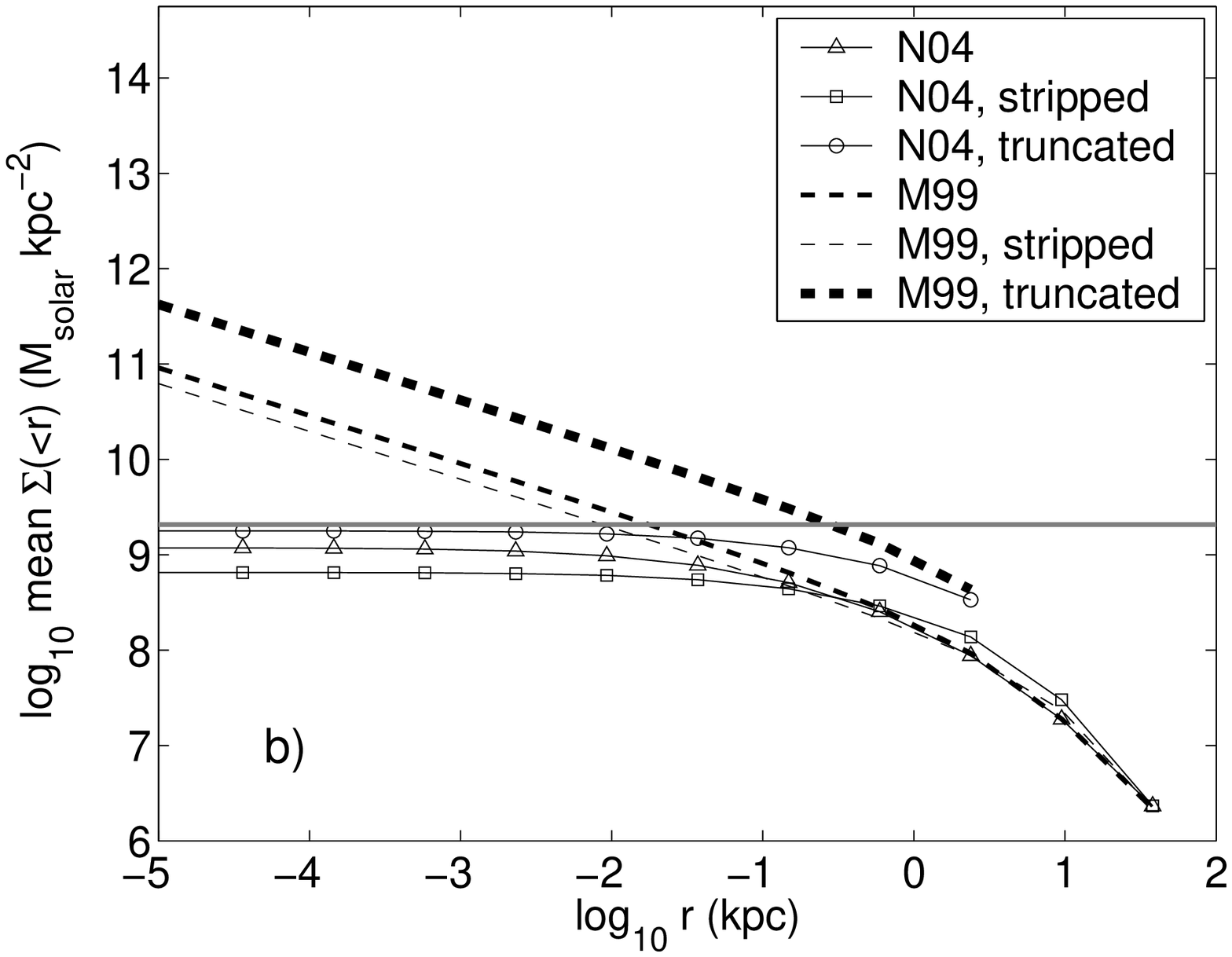}
\caption{{\bf a)} Mean surface mass density $\bar{\Sigma}(<r)$ profiles for a $10^{10} \ M_\odot$ subhalo at $z_\mathrm{l}=0.5$: SIS (thin solid), truncated SIS (thick solid), NFW (thin dash-dotted), truncated NFW (thick dash-dotted), H03 (thick dotted) and K04 (thin dotted). {\bf b)} Same for N04 (solid with triangles), stripped N04 (solid with squares), truncated N04 (solid with cirles), M99 (medium dashed), stripped M99 (thin dashed) and truncated M99 (thick dashed) profiles. In both panels, the horizontal gray line indicates the critical surface mass density for strong lensing, $\Sigma_\mathrm{c}$. The truncated models have been constructed from density profiles for halos of mass $10^{10}\ m_\mathrm{bnd}^{-1} \ M_\odot$ (with $m_\mathrm{bnd}=0.03$), abruptly terminated at a radius enclosing a mass of $10^{10}\ M_\odot$. The stripped versions of the N04 and M99 profiles have been constructed using (\ref{stripeq}). 
\label{fig1}}
\end{figure*} 

Based on similar simulations at higher resolution, \citet[][hereafter K04]{Kazantzidis et al.} et al. instead find a density profile of the form: 
\begin{equation}
\rho_\mathrm{K04}(r)=kr^{-\gamma}\exp(-\frac{r}{r_\mathrm{b}}),
\end{equation}
where $\gamma\approx 1\pm0.05$, $0.1\lesssim r_\mathrm{b}/r_\mathrm{S,NFW}\lesssim 1.2$ and $r_\mathrm{S,NFW}$ refers to the corresponding parameter of the progenitor halo, which we assign a mass analogous to the H03 case. Here, $k$ is a normalization factor which we introduce to scale the subhalo to its current mass $m_\mathrm{bnd}M_\mathrm{prog}$. In the following, we adopt $\gamma=1$ and $r_\mathrm{b}/r_\mathrm{S,NFW}=0.75$.

In the case of both the H03 and K04 profile , we assume the subhalo mass to be entirely contained within $r_\mathrm{vir}$ of its progenitor and adopt $m_\mathrm{bnd}=0.03$ as a typical value for subhalos of galaxy-mass halos \citep{van den Bosch et al. b}. 

While many previous investigations of the lensing effects produced by subhalos have assumed a sharp density cutoff at some outer radius, it is clear from the H03 and K04 profiles that the density decrease in the outer regions is expected to be gradual. To demonstrate the effects of this, we consider two different modifications of the M99 and N04 profiles -- one with a sharp truncation radius $r_\mathrm{t}$ such that the mass contained inside becomes $M_\mathrm{sub}(<r_\mathrm{t})=m_\mathrm{bnd}M_\mathrm{prog}$, and one in which the halo is gradually stripped of material according to the function proposed by H03: 
\begin{equation}
\rho_\mathrm{M99/N04,stripped}(r)=\frac{f_\mathrm{t}}{1+(r/r_\mathrm{te})^3}\rho_\mathrm{M99/N04}(r).
\label{stripeq}
\end{equation}
 Lacking a better prescription, we adopt the $f_\mathrm{t}(m_\mathrm{bnd})$ and $r_\mathrm{te}(m_\mathrm{bnd})$ relations presented in H03, although these will in practice depend on the details of the progenitor halo model assumed.

When considering the impact of stripping and/or truncation on lensing properties, we have chosen to compare halos of fixed mass, which is defined as the virial mass for unstripped halos, and as the total subhalo mass after stripping for the subhalos. This choice -- which reflects the tradition in the field of subhalo simulations, where one usually present the mass function of the subhalos themselves and not their progenitors -- has important consequences.  For instance, a sharply truncated halo of a certain mass has a much smaller spatial extent than an untruncated halo of the same mass.  Consequently, the density in the inner regions of the truncated halo is necessarily higher than that in the untruncated halo of the same mass, and therefore, the truncated halo will have a larger Einstein radius. However, a more gradual density decrease in the outskirts can in principle drive the Einstein radius in either direction, depending on the details of the stripping process. As it turns out, the stripping procedure adopted here (equations (\ref{H03eq}) and (\ref{stripeq})) tends to produce Einstein radii that are smaller than those of the unstripped profiles.

\section{Strong-lensing image separation versus observational angular resolution}
To first order, the image separation produced by an extended object with a density that decreases as a function of distance from the centre is given by 
\begin{equation}
\Delta\theta\approx 2 R_\mathrm{E}/D_\mathrm{ol}, 
\label{imagesepeq}
\end{equation}
where $D_\mathrm{ol}$ represents the angular-size distance between observer and lens, and  $R_\mathrm{E}$ represents the linear Einstein radius. The latter is defined as the radius inside which the mean surface mass density $\bar{\Sigma}$ of the lens equals the critical surface mass density 
\begin{equation}
\bar{\Sigma}(<R_\mathrm{E})=\Sigma_\mathrm{c}=\frac{c^2D_\mathrm{os}}{4\pi G D_\mathrm{ol}D_\mathrm{ls}},
\end{equation}
where $D_\mathrm{os}$ and $D_\mathrm{ls}$ are the angular-size distances between observer and source, and lens and source, respectively.
 
From the density profiles described in the previous section, we numerically compute the mean surface mass density profiles $\bar{\Sigma}(<R)$, shown in Fig.~\ref{fig1}a and b in the case of a subhalo lens mass of $10^{10} M_\odot$. At the adopted lens redshift, the smallest linear radius plotted corresponds to an angular Einstein radius of $\approx 3\times 10^{-6}$ arcseconds, which is below the angular resolution of any telescope that will become operational in the foreseeable future. 

In fact, only the SIS and M99 profiles cross the $\Sigma_\mathrm{c}$ divide (gray horizontal line) at radii substantially larger than this, which means that only these profiles will give rise to detectable image separations for a subhalo of this mass. In its original form, the NFW profile gives $\bar{\Sigma}(<R)$ lower than $\Sigma_\mathrm{c}$ at these radii, whereas the truncated version leads to $\bar{\Sigma}(<R)$ above the threshold. Both the H03 and K04 subhalo models, which should be far more realistic than any sharp truncation, do however predict $\bar{\Sigma}(R)$ smaller than that of an undisturbed NFW halo of the same mass. The N04 profile (in both its original, stripped and truncated form) also ends up below the threshold, due to its finite-density core. The effects of truncation versus gradual stripping are qualitatively similar for the NFW, N04 and M99 halos -- a sharp truncation increases the Einstein radius whereas a gradual stripping decreases it relative to the Einstein radius produced by the original profile. This indicates that previous investigations based on sharp outer truncations are likely to have {\it substantially overestimated the image separations.} 
\begin{figure*}
\plottwo{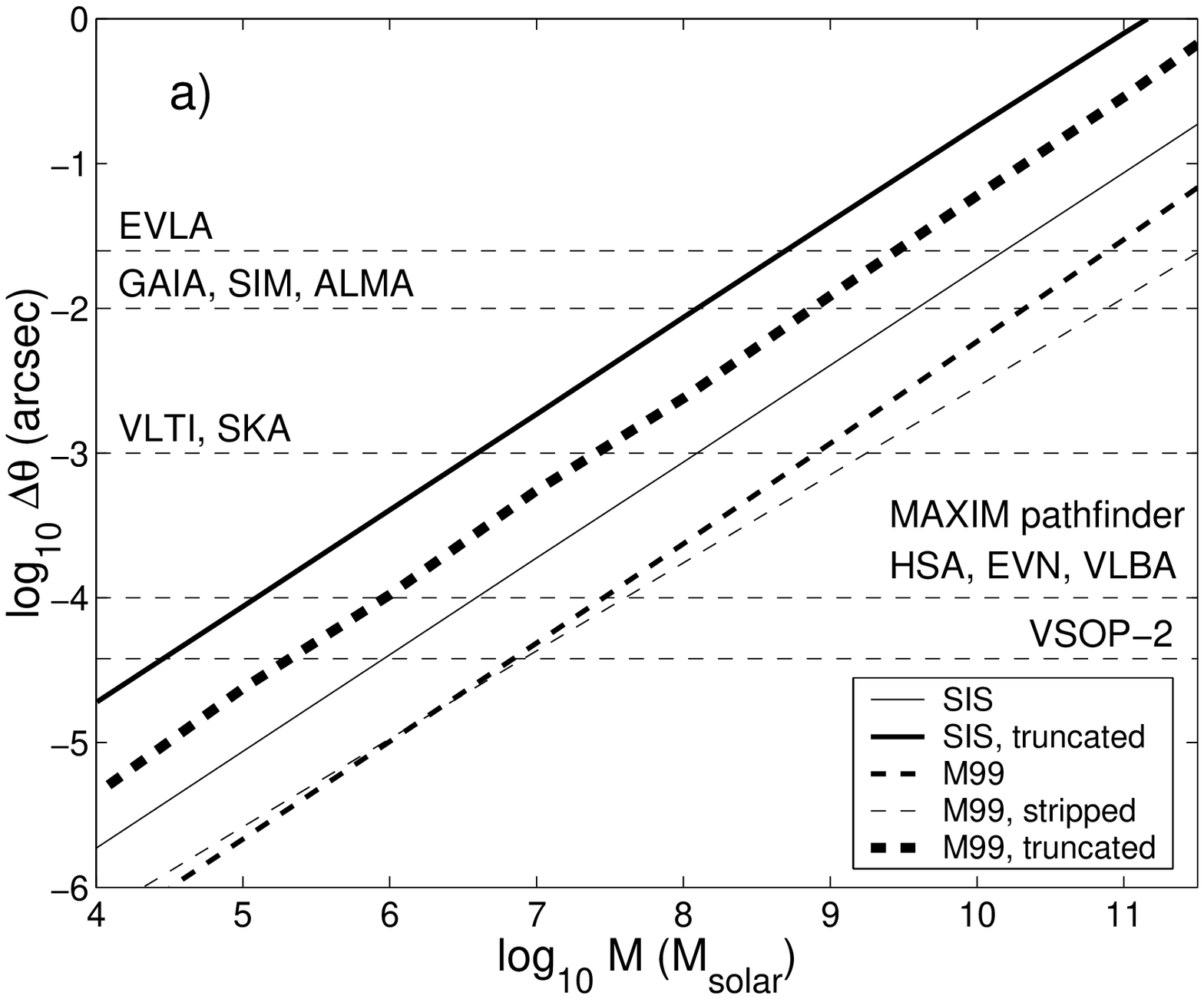}{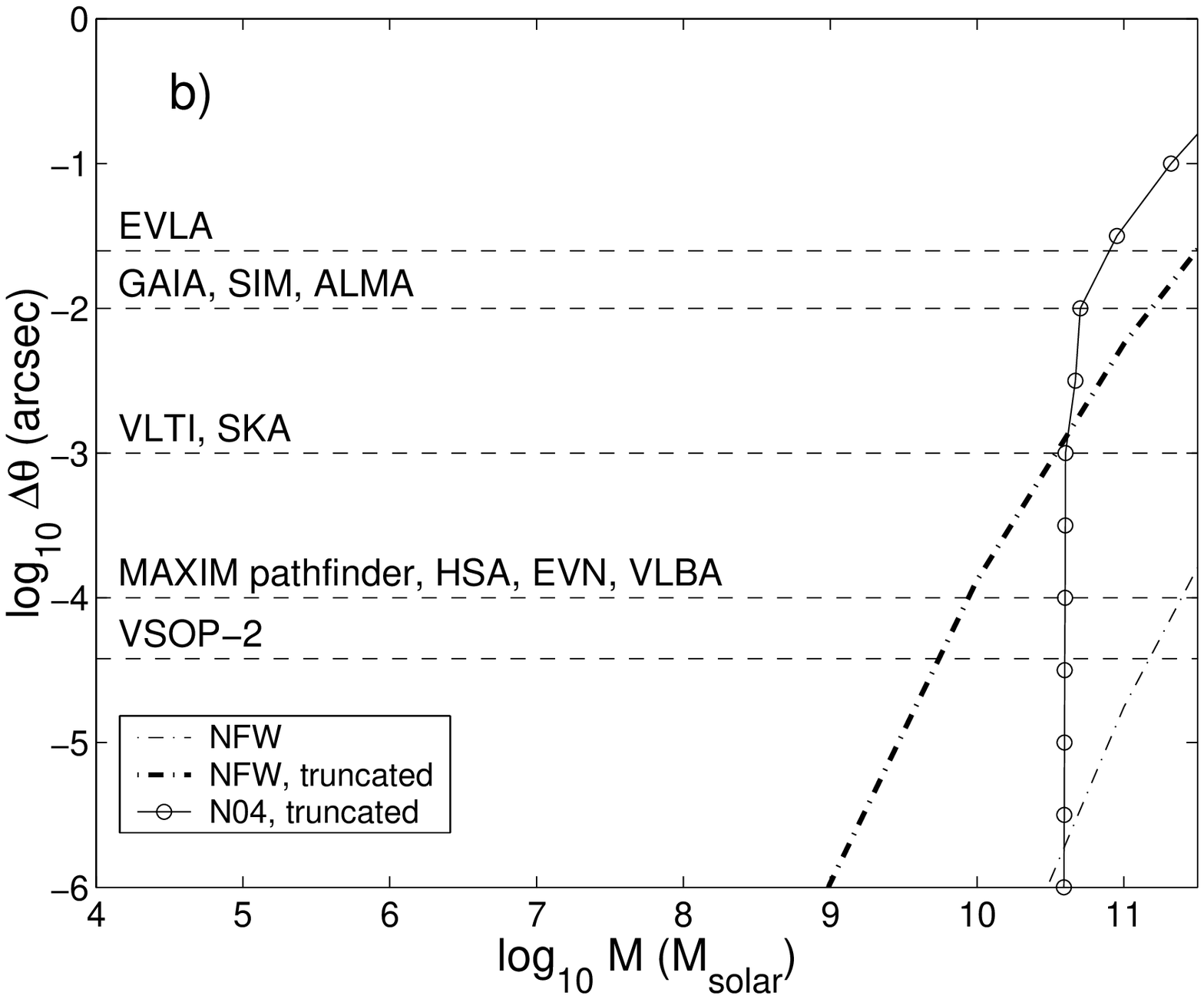}
\caption{Subhalo mass versus image separation $\Delta\theta$ for those density profiles from  Fig~\ref{fig1} that give rise to image separations on scales larger than microarcseconds. The angular resolution of a number of existing and planned observational facilities have been indicated by horizontal dashed lines, marked with labels (see main text for details). {\bf a)} The different diagonal lines represent SIS (thin solid), truncated SIS (thick solid), M99 (medium dashed), stripped M99 (thin dashed) and truncated M99 (thick dashed) subhalo models. {\bf b)} The different diagonal lines represent NFW (thin dash-dotted), truncated NFW (thick dash-dotted) and truncated N04 (solid with circles) subhalo models. 
  \label{fig2}}
\end{figure*}

In Fig.~\ref{fig2}, we plot the image separations predicted for $10^4$--$10^{11}\ M_\odot$ subhalos against the angular resolution of a number of planned or existing observational facilities, operating at a wide range of wavelengths. These include the proposed MAXIM pathfinder in X-rays\footnote{maxim.gsfc.nasa.gov/pathfinder.html}; VLTI with the proposed VSI instrument \citep{Malbet et al.}, the planned GAIA\footnote{www.rssd.esa.int/index.php?project=GAIA} and SIM PlanetQuest\footnote{planetquest.jpl.nasa.gov/SIM/} satellites in the optical/near-infrared; the currently available EVN\footnote{www.evlbi.org}, HSA\footnote{www.nrao.edu/HSA/}, VLBA\footnote{www.vlba.nrao.edu} arrays plus the planned ALMA\footnote{www.alma.info}, EVLA\footnote{www.aoc.nrao.edu/evla/}, SKA\footnote{www.skatelescope.org} arrays, and also space-VLBI with the planned VSOP-2\footnote{www.vsop.isas.ac.jp/vsop2/} satellite at radio wavelengths. Please note that here we consider only the best resolution limits attainable with these telescopes, whereas the resolution at the wavelengths that maximize the number of observable high-redshift sources may be considerably worse. 

It is immediately evident from Fig.~\ref{fig2} that there are large differences between the image separation predictions of the various halo models. As the discrepancy between the number densities of luminous galaxies and dark matter halos does not start to become severe until the halo mass drops below $\lesssim 10^{10} \ M_\odot$ \citep[e.g.][]{Verde et al., van den Bosch et al. a}, subhalos at masses below this limit need to produce measurable image separations ($\theta \gtrsim 4\times 10^{-5}$ arcsec for VSOP-2, which has the best theoretical resolution among the telescopes included in Fig. ~\ref{fig2}) in order for dark galaxies to be detectable through image-splitting effects. Out of the halo models tested, only two actually meet this criterion without adhering to sharp truncations: the SIS and the M99 halos. The H03 and K04 profiles both give image separations smaller than $10^{-6}$ arcsec for all the halo masses considered and are therefore completely outside the plotted region. Even in the optimistic case of an M99 halo (in either its original or stripped form, whereas the sharp truncation, as discussed previously, is not considered realistic), the image separations are a factor of $\approx 3$--7 smaller than those predicted for a SIS (and $\approx30$--60 times smaller than those of a truncated SIS), rendering only the few most massive subhalos ($\sim 10^{10} \ M_\odot$ or slightly higher) detectable at $\sim 0.01\arcsec$ resolution (GAIA, SIM and ALMA). At milliarcsecond resolution (VLTI and SKA), dark galaxies with masses $\gtrsim10^9 \ M_\odot$ may become detectable. To probe further down the subhalo mass function, submilliarcsecond-resolution facilities (MAXIM pathfinder, HSA, EVN, VLBA or VSOP-2) will be required. 

These image separations have been computed for fiducial lens and source redshifts of $z_\mathrm{l}=0.5$ and $z_\mathrm{s}=2.0$, but the overall picture does not change substantially for other realistic choices of these parameters. A higher $z_\mathrm{s}$ implies a lower critical surface mass density $\Sigma_\mathrm{c}$ for certain $z_\mathrm{l}$, which for a fixed subhalo density profile will increase the Einstein radius and boost the image separation $\Delta\theta$. This boost becomes more pronounced for objects with flat inner $\bar{\Sigma}(<R)$ profiles, like the NFW, N04, H03 and K04 density profiles (see Fig.~\ref{fig1}). While the density profiles themselves also change with $z_\mathrm{l}$ through the redshift evolution of the virial radius and the concentration parameter, this has a minor effect on $\Delta\theta$, compared to impact of the $\Sigma_\mathrm{c}(z_\mathrm{l},z_\mathrm{s})$ dependence. We find, that by allowing source redshifts as high as $z_\mathrm{s}=5.0$, the image separation of a $10^{10}\ M_\odot$ NFW-type subhalo can be increased by a factor of up to $\approx 30$. Even though this boost factor may seem large, the resulting image separation is still only $\Delta\theta\sim 10^{-6}$ arcseconds, and hence far below the observable range. For an M99 profile, the corresponding boost is only $\approx 30\%$. Hence, subhalos need to have steep inner density profiles (of approximately the M99 type) to become observable, even if a sample of multiply-imaged quasars with lens and source redshifts chosen to maximize the subhalo image separations would become available.  

\section{External convergence and shear}
The $\Delta\theta$ estimates presented in the previous section are based on the assumption that the subhalos can be treated as isolated objects. The effects of external convergence $\kappa$ and shear $\gamma$ due to the galaxy (and halo) hosting these subhalos have thereby been neglected. \citet{Yonehara et al.} explored the consequeces of non-zero $\kappa$ and $\gamma$ in the case of SIS subhalos and found the external potential to have a non-negligible impact on the strong lensing properties of such objects. However, their study suggests that the resulting image separations $\Delta\theta_{\kappa,\gamma}$ are typically within a factor of  $\approx 3$ from those derived in the case of $\kappa=\gamma=0$. If applied to the predictions plotted in Fig.~\ref{fig2}, a boost factor of this magnitude would be insufficient to challenge our main results, namely that only subhalos with inner density profiles as steep as those of M99 (or SIS) models produce image separations that can be resolved with current or planned telescopes. The question remains whether the boost factor $f_\mathrm{boost}=\Delta\theta_{\kappa,\gamma}/\Delta\theta_{\kappa=0,\gamma=0}$ can become considerably larger for some of the more realistic density profiles considered here. 

To investigate this, we use ray-tracing simulations to numerically assess the distribution of $f_\mathrm{boost}$. For the macrolens (i.e. the halo and galaxy hosting the subhalo) we adopt an SIS density profile with  $\sigma_v=150$ km s$^{-1}$. At $z_\mathrm{l}=0.5$ and $z_\mathrm{s}=2.0$, the corresponds to a linear (angular) Einstein radius of 2.5 kpc (0.4 arcsec). Subhalos are then distributed within this structure, assuming an NFW profile with $c=10$ for the subhalo component of the CDM in the macrolens. Since the optical depth for image splitting by subhalos is low \citep{Yonehara et al.}, the $f_\mathrm{boost}$ distribution is not very sensitive to the exact spatial distribution of the subhalos within the host halo. For simplicity, we assume all subhalos to have the same mass in each simulation, but repeat the simulations for different subhalo masses to explore the dependence of $f_\mathrm{boost}$ on the subhalo mass. The sources are assumed to be point-like and are distributed on a regular grid in the source plane. To assess the effects of magnification bias, two different cases are considered: one in which no magnification threshold is imposed, and one in which only source positions which produce total magnifications $\mu \geq 10$ are analyzed. Source positions for which multiple images are not produced are always rejected.

The vast majority of the resulting macroimages turn out to be unaffected by the subhalos due to the low optical depth, and are therefore discarded. The properties of the Einstein ring of the subhalo is calculated for the remaining macroimages, and the image separation $\Delta\theta_{\kappa,\gamma}$ estimated from its angular diameter. In the case of $\gamma\ne0$, the Einstein ring becomes an ellipse, and $\Delta\theta_{\kappa,\gamma}$ is then estimated along the major axis. This leads to a systematic overestimate of $\Delta\theta_{\kappa,\gamma}$, which ensures that the resulting boost factors are conservative upper limits.
\begin{figure}
\plotone{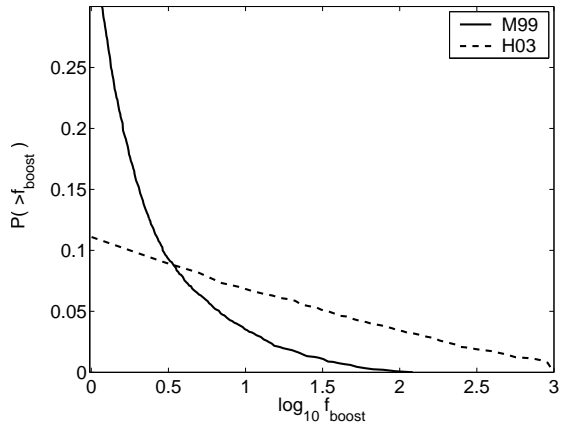}
\caption{The cumulative probability distribution $P(>f_\mathrm{boost}$) of image separation boost factors $f_\mathrm{boost}$ due to the external potential in the case of M99 (solid) and H03 (dashed) subhalo density profiles. In both cases, the mass of the subhalo is $10^{10}\ M_\odot$.\label{fig3}}
\end{figure}

In the case of no magnification threshold, the resulting distribution of $f_\mathrm{boost}$ is shown in Fig.~\ref{fig3} for an M99 subhalo (solid line) and an H03 subhalo (dashed), both with $M_\mathrm{sub}=10^{10} \ M_\odot$. As discussed in Section 3, these subhalo models span the range from effectively undetectable (H03) to favourable for detection (M99). In both cases, the tail of the $f_\mathrm{boost}$ distribution extends up to very high values, which means that it is in principle possible to find macroimages for which subhalos of masses much lower than indicated by Fig.~\ref{fig2} can be detected through image-splitting effects. However, such macroimages are exceedingly rare. The probability of having $f_\mathrm{boost}\geq 100$ is $\sim 10^{-3}$ in the case of M99 and $\approx 0.03$ in the case of H03. The expectation value for the boost factor is $\langle f_\mathrm{boost}\rangle\approx 2.3$ for the M99 subhalo and $\langle f_\mathrm{boost}\rangle\approx 14$ for the H03 subhalo. This is enough to shift the $\Delta\theta$ prediction for the M99 subhalos fairly close to that of an untruncated SIS in Fig.~\ref{fig2}, but insufficient to move the H03 prediction ($\Delta\theta_{\kappa=0,\gamma=0} \sim 10^{-15} $ arcsec at this mass) into the detectable range. In the case of M99 subhalos, we have also explored the mass dependence of $\langle f_\mathrm{boost}\rangle$, finding $\langle f_\mathrm{boost}\rangle\approx 3.1$ for $M_\mathrm{sub}=10^8 \ M_\odot$ and $\langle f_\mathrm{boost}\rangle\approx 3.6$  for $M_\mathrm{sub}=10^6 \ M_\odot$. This trend with increasing $\langle f_\mathrm{boost}\rangle$ for decreasing subhalo mass translates into a slight change in slope for the $\Delta\theta(M_\mathrm{sub})$ prediction in Fig.~\ref{fig2}, but has no dramatic effects on our overall conclusions. 

Flux-limited observations are affected by a magnification bias, since flux threshold of the survey tends to favour highly magnified quasars from the vast population of intrinsically faint objects over modestly magnified (or even demagnified) quasars from the relatively small population of intrinsically bright objects. The exact impact of this effect depends on the flux threshold and the quasar luminosity function (which, in turn, depends on the wavelength at which the quasars are observed), and is outside the scope of this paper. Here, we instead make a crude assessment of the likely impact of magnification bias by rejecting all systems which do not fulfill the criterion $\mu \geq 10$. As expected, this increases the boost of the image separation produced by the subhalos. In the case of $M_\mathrm{sub}=10^{10} \ M_\odot$ subhalos, the boost reaches $\langle f_\mathrm{boost}\rangle\approx 6.7$ for an M99 profile and $\langle f_\mathrm{boost}\rangle\approx 85$ for an H03 profile, i.e. an increase by a factor of $\approx 3$ (M99) and $\approx 6$ (H03) compared to the case without magnification bias. While this brightens the prospects of detecting subhalos with M99 profiles, H03 subhalos still produce image separations that are too small to be resolved with any of the telescopes considered in Fig.~\ref{fig2}.  

\section{Discussion}
Our results indicate that detection of subhalos through gravitational image-splitting is likely to be considerably more challenging than suggested in previous studies, due to the smaller image separations predicted for subhalo density profiles more realistic than the SIS models often adopted. In fact, no currently planned telescope will be able to resolve the image separations produced by subhalos with density profiles of the type suggested by the most realistic simulations currently available (H03 \& K04). We stress, however, that these simulations do not necessarily represent the final word on this issue. NFW density profiles were adopted for the subhalo progenitors in both the H03 and K04 simulations, which means that the slope of the central density profile prior to stripping was {\it assumed} rather than derived from the simulations themselves. If the subhalos would have steeper central density slopes (e.g. of M99 type), these would give rise to image separations that could be resolved even with existing telescopes. 

Despite the somewhat bleak detection prospects presented here, there are at least two effects that can potentially improve the detectability of image splitting by subhalos: baryon cooling and the presence of intermediate mass black holes. 

Baryon cooling would cause the subhalo (or its progenitor halo) to contract \citep[e.g.][]{Gnedin et al.,Maccio et al.,Gustafsson et al.,Kampakoglou}, thereby increasing the central density and boosting the image separation. It is, however, not clear how strong this effect is likely to be, since this depends on the details of the mechanism that prevents the dark subhalo from forming stars. Baryon cooling is usually assumed to be associated with star formation, and most attempts to explain the lack of bright subhalos therefore propose that the baryons have either been lost early in the history of the Universe, or are prevented from cooling by the ultraviolet background provided by reionization \citep[e.g.][]{Barkana & Loeb,Read et al.}. Neither mechanism is likely to result in any significant halo contraction. Nonetheless, claims of dark baryons in the form of cold gas in galactic disks have been made \citep{Bournaud et al.}, implying that there may be some route for gas to cool without forming stars or being detected by the usual H$_2$ tracers \citep[e.g.][]{Pfenniger et al.,Pfenniger & Combes}. One may therefore speculate that there could be alternative solutions to the missing-satellite problem, in which subhalos are kept dark even though baryon cooling has taken place.

Intermediate mass black holes \citep[IMBHs, see e.g.][]{van der Marel,Zhao & Silk,Noyola et al.} with masses of $\sim 10^2$--$10^4\ M_\odot$ would also boost the image separations, if present in the centres of dark subhalos. This could for instance be the case if the empirical relations between the mass of a supermassive black hole and its dark matter halo \citep{Ferrarese} would extend into the dwarf-galaxy mass range. Even in the case of $\kappa=\gamma=0$, an IMBH of mass $10^4 \ M_\odot$, would give an image separation of $\approx 4\times 10^{-4}$ arcsec for $z_\mathrm{l}=0.5$ and $z_\mathrm{s}=2.0$. This is sufficient to allow VSOP-2 to resolve the image splitting, regardless of the density profile of the subhalo hosting the IMBH.  Of course, even if the Ferrarese relations would hold for luminous dwarf galaxies, the do not necessarily do so for dark ones, since this depends on the formation details of IMBHs. Moreover, a population of halo IMBHs {\it not} associated with subhalos could form an undesired background of millilensing events that would obfuscate attempts to study subhalos through image-splitting effects.

Finally, there is an important issue related to the limitation of current CDM halo simulations. Current N-body simulations only resolve scales down to $R\sim 0.001 r_\mathrm{vir}$, whereas for the steeper density profiles -- i.e. those producing detectable image separations for dwarf-galaxy masses -- a non-negligible fraction of $\bar{\Sigma}(<r)$ at the  Einstein radius comes from radii smaller than this. This makes the exercise of deriving dark halo lensing properties from fitting functions based on current N-body data somewhat hazardous. Unless theoretical arguments \citep[e.g.][]{Hansen & Stadel} can be used to determine the slope of the density profile at even smaller radii, robust predictions for the image separations produced by low-mass halos may be very difficult to derive. Observational limits on the lensing optical depth as a function of $\Delta\theta$, based on searches for subhalo image splitting, may eventually turn out to be a useful way of setting constraints on the internal structure of dark matter halos at scales below the resolution of numerical simulations. Of course, the optical depth does not depend only on the density profile of the subhalos, but also on their mass function and the spatial distribution within the macrolens. Detailed estimates of the optical depth for strong lensing by subhalos in the dwarf-galaxy mass range is outside the scope of this study, but will be covered in paper II of this series (Riehm et al. 2008, in preparation).

\acknowledgments{EZ acknowledges research grants from the Swedish Research Council, the Royal Swedish Academy of Sciences and the Academy of Finland. The anonymous referee is thanked for insightful comments which helped improve the quality of the manuscript.}

\end{document}